# Phase nucleation, coarsening and evolution pathways of a sputtered nanostructured Inconel 725 alloy during heat treatment

Ruqing Cao [1], Ikponmwosa J. Iyinbor [2], Andrea M. Hodge [2,3], Timothy J. Rupert [1,4,*]

[1] Hopkins Extreme Materials Institute, Johns Hopkins University, Baltimore, MD 21218, USA

[2] Mork Family Department of Chemical Engineering and Materials Science, University of Southern California, Los Angeles, CA 90089, United States

[3] Department of Aerospace and Mechanical Engineering, University of Southern California, Los Angeles, CA 90089, United States

[4] Department of Materials Science and Engineering, Johns Hopkins University, Baltimore, MD 21218, USA

[*] Corresponding Author: tim.rupert@jhu.edu

## Abstract

Physical vapor deposition enables the fabrication of nanostructured superalloys with unique defect architectures, yet their phase evolution pathways can differ significantly from those of conventionally processed alloys. In this study, the effects of solution and aging treatments on phase selection and precipitation behavior in sputtered Inconel 725 films with an initially uniform columnar nanotwinned structure were systematically investigated. Direct aging at relatively low temperatures promoted extensive δ-phase precipitation at twin boundaries and defect-rich regions, which depleted Nb from the γ matrix and suppressed γ′/γ″ precipitation. In contrast, high-temperature solution treatment induced recrystallization and eliminated the nanotwinned structure, significantly reducing δ-phase precipitation and increasing Nb availability to enable the formation of ultrafine spherical $\gamma'/\gamma''$ precipitates within a refined $\gamma$ matrix (<1 μm). Subsequent aging treatments promoted elemental partitioning and drove the morphological evolution of $\gamma'/\gamma''$ precipitates from spherical to lenticular forms, while $\delta$ precipitation became increasingly concentrated along grain boundaries. This spatial separation of intragranular $\gamma'/\gamma''$ and grain boundary $\delta$ phases enabled simultaneous precipitation strengthening and grain stabilization, resulting in hardness values approaching 9 GPa. As a whole, this study demonstrates that the initial templates provided by defect structures can govern phase selection and precipitation pathways, providing a strategy for tailoring microstructure and achieving synergistic strengthening in nanostructured superalloys.

## 1. Introduction

Nanotwinned structures in metals and alloys can exhibit high yield strength, ductility [1-4], and thermal stability [5-8], which are often difficult to achieve simultaneously in conventional nanograined materials. Coherent twin boundary energies are an order of magnitude lower than those of most high-angle grain boundaries in metals and alloys such as Cu [7], Ni [9], Al-Fe [10], and Ni-Cr [11], which can significantly increase the coarsening temperature, indicating the superior stability of twinned nanostructures. However, existing studies on twinned structures have predominantly focused on their resistance to grain growth, while the effects of twin boundaries on phase nucleation and phase evolution during thermal treatments remain far less explored.

Recently, Bahena et al. reported a unique nanotwinned microstructure in a 15-μm-thick sputtered Inconel 725 film with a highly uniform elemental distribution [12]. With the strong precipitation tendency of Inconel 725 alloy during thermal treatment [13-18], understanding phase evolution in this metastable nanotwinned alloy is particularly important for tailoring mechanical performance. Hodge and coworkers further demonstrated that aging at 730 °C for 8 h induced microstructural evolution accompanied by extensive precipitation of δ phases and metallic carbides [12, 19]. Notably, γ′ (primitive cubic $Ni_3Al$) and γ″ (body-centered tetragonal $Ni_3Nb$) precipitates, which typically precipitate during aging and serve as the primary strengthening phases in Inconel 718 and 725 alloys [13, 18, 20, 21], were found to be absent in the sputtered films [12, 19]. This observation raises fundamental questions regarding the precipitation behavior in sputtered nanotwinned Inconel 725 and its departure from that of conventional wrought and cast alloy counterparts. Specifically, the competitive precipitation mechanism, in which early δ-phase formation at twin boundaries and associated solute partitioning impedes the nucleation of γ′/γ″ precipitates.

A critical challenge is therefore to identify the heat-treatment pathways and microstructural conditions necessary to promote γ′/γ″ precipitation in sputtered nanotwinned Inconel 725 alloys. An important question is whether this precipitation pathway can be altered through an intermediate solution treatment. In conventional wrought and cast Inconel alloys, a solution treatment followed by aging is the standard pathway for achieving controlled γ′/γ″ precipitation, as the solutionizing step dissolves pre-existing precipitates and homogenizes elemental segregation to create favorable conditions for subsequent γ′/γ″ precipitation during aging [20-22]. However, the role of such a solution treatment in sputtered nanotwinned Inconel 725 films remains unclear because these films already exhibit a chemically uniform microstructure yet contain unique defect structures and residual stresses introduced during deposition. In particular, it is not clear if a solution treatment fundamentally alters the defect-mediated precipitation pathway in the as-sputtered films.

In this work, we systematically investigate how different combinations of solution and aging treatments influence phase evolution in sputtered Inconel 725 films. Particular attention is given to the competitive formation of γ′, γ″, δ phases, and Cr-rich precipitates, as well as the relationship between phase selection, recrystallization behavior, and hardness. By comparing direct aging with aging following solution treatment, this study aims to clarify how the microstructure prior to aging governs subsequent phase evolution. More importantly, this work demonstrates that precipitation pathways can be tailored through defect network reconstruction, enabling a spatial decoupling of competing phases and providing a generalizable strategy for achieving synergistic precipitation strengthening in nanostructured superalloys.

## 2. Experimental

### 2.1 Sample preparation

Sputtered coatings were deposited using a 7.62 cm diameter Inconel 725 planar target onto a 5.08 cm diameter sapphire C-plane substrate via direct current (DC) magnetron sputtering. The deposition was carried out at a 5.08 cm target–substrate distance with a power of 1500 W, an Ar working pressure of 0.3 Pa, a chamber base pressure of $9.3 \times 10^{-4}$ Pa, and a sputtering rate of 9.2 nm/s. The resulting films were ~13.5 μm thick, as determined using an AMBIOS XP-2 Profilometer. Top surface compositional data for the as-sputtered film were measured to be 52.6 Ni, 24.0 Cr, 8.1 Mo, 10.6 Fe, 3.3 Nb, 1.1 Ti, and 0.3 Al wt.% using the energy-dispersive x-ray spectroscopy (EDS) capabilities of a Helios G4 P-FIB UXe-DualBeam microscope at 20 kV. Prior to heat treatment, the films were peeled from the substrate to obtain free-standing samples. Individual pieces were then subjected to varying combinations of solution and aging treatment in a GSL-1100X tube furnace (MTI Corporation). The overall heat-treatment schedule is illustrated in Fig. 1, with the most complex pathway shown as a representative example. A solutionizing heat-treatment step was performed at 1038 °C for 2 h followed by air cooling, and the sample subjected solely to this treatment is hereafter referred to as the *Solutionized sample*. This name is kept to remain consistent with literature reports for processing of traditional Inconel alloys, although we will show that its purpose is in fact not to solutionized the specimen in these nanotwinned films. After the solution treatment, the aging process followed two routes. The single-aging treatment was conducted at 730 °C for 8 h, followed by furnace cooling; the corresponding sample is denoted as the *Solutionized + single-aged sample*. The dual-aging treatment involved sequential annealing at 730 °C and 620 °C for 8 h each, with furnace cooling between the aging steps and then subsequent air cooling to room temperature; this condition is denoted as the *Solutionized + dual-aged sample*. In addition, two comparative samples that were subjected to only single- and dual-aging without prior solutionizing treatment were also prepared. These are referred to as the *Single-aged sample* and the *Dual-aged sample*, respectively.

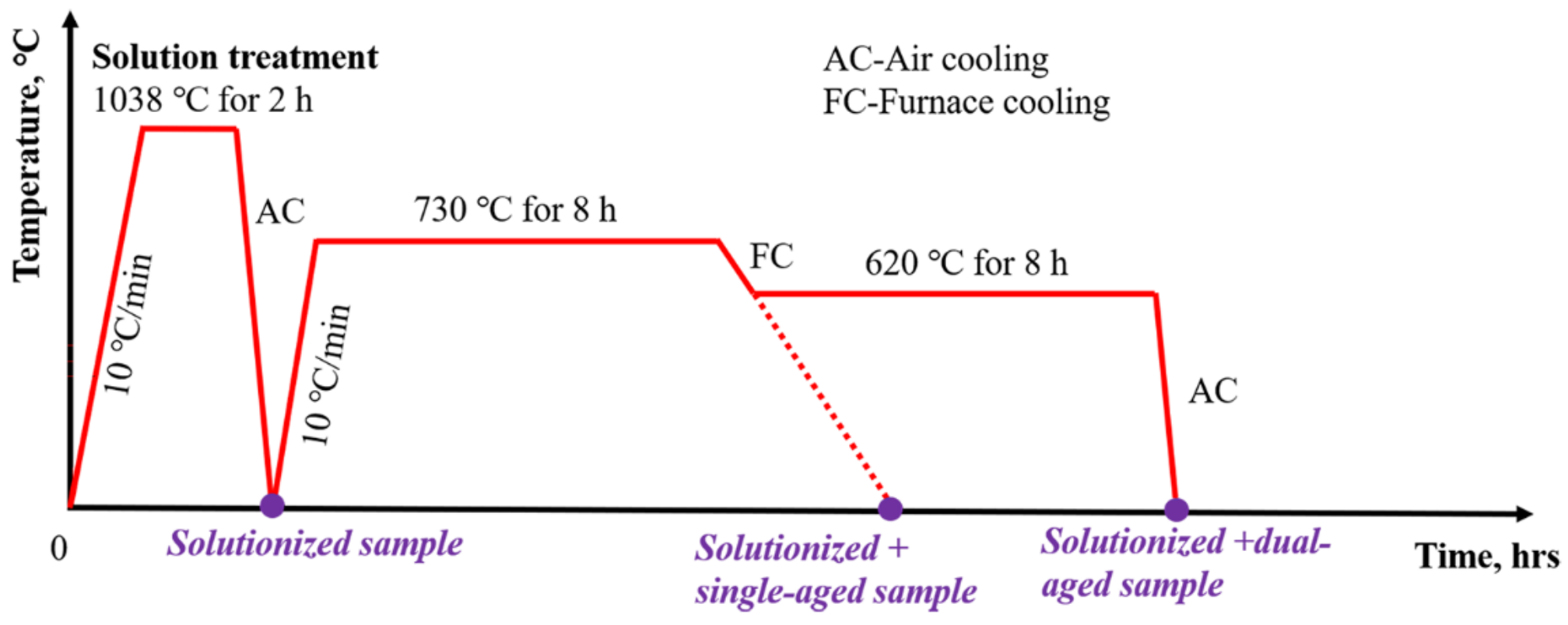

**Fig. 1. The heat treatment procedures for the samples include: *Solutionized sample*, *Solutionized + single-aged sample*, and *Solutionized + dual-aged sample*. Comparison *Single-aged* and *Dual-aged samples*, without solutionization treatment, were also created.**

## 2.2 Microstructure Characterization

Cross-sectional micrographs, and precipitation behavior of the δ phase and carbides under different heat treatments, were investigated using transmission electron microscopy (TEM) methods, including scanning transmission electron microscopy (STEM) and energy-dispersive X-ray spectroscopy (EDS). The analyses were conducted on a JEOL JEM-F200 TEM equipped with a field emission gun (FEG) and operated at 200 kV, with EDS measurements performed using dual silicon drift detectors (SDDs). TEM lamellae were prepared by the focused ion beam (FIB) lift-out method using a Thermo Fisher Helios G5 scanning electron microscope (SEM)-FIB. To minimize $Ga^+$ beam induced damage on the film surface, a ~2 μm-thick platinum protective layer was deposited on the region of interest prior to milling. A combination of bright-field (BF) imaging, dark-field (DF) imaging, center dark-field (CDF) imaging, and selected-area electron diffraction (SAED) techniques were employed to characterize γ′/γ″ precipitates within the γ matrix. In the *As-sputtered* sample, columnar grain widths were measured from BF-TEM images, whereas twin thicknesses were measured from high-resolution TEM images. In addition, STEM and EDS analyses were performed in the γ phase.

### 2.3 Nanoindentation

Nanoindentation measurements were conducted using a Hysitron TriboIndenter equipped with a diamond Berkovich tip with a nominal tip radius of 100 nm. The tip area function was calibrated against a fused quartz standard prior to testing using 100 indents, employing a loading profile of 5 s loading, 2 s hold, and 5 s unloading, with maximum loads ranging from 100 to 10,000 μN. A 7 × 7 array of indents, spaced 10 μm apart, was performed on all samples under force-controlled loading at a constant loading rate. Applied loads ranged from 7000 to 8000 μN, resulting in maximum indentation depths of approximately 230 nm, corresponding to less than 10% of the film thickness.

## 3. Results and Discussion

### 3.1 As-deposited nanotwinned microstructure

Fig. 2A presents a BF-STEM image of the cross-sectional morphology of the as-sputtered Inconel 725 sample, revealing a uniform columnar structure that extends through the film thickness. A higher-magnification BF-STEM image (Fig. 2B) together with corresponding EDS elemental maps of Ni, Cr, and Nb confirms a uniform elemental distribution, indicating an initially homogeneous chemical composition despite the presence of numerous interfacial defects. This behavior is consistent with a typical sputtering process, where rapid cooling rates limit diffusion and prevent segregation and formation of equilibrium phases, often resulting in a metastable solid solution [23, 24]. As shown in Figs. 2C and 2D, the as-deposited film consists of extremely fine nanotwins oriented perpendicular to the film growth direction, with twin boundaries clearly identified in high-resolution TEM image (marked by dashed yellow lines). Statistical analysis (Fig. 2E) shows an average columnar grain width of 28.7 ± 7.3 nm and an average twin thickness of 0.6 ± 0.3 nm. These results confirm that the as-sputtered film possesses a chemically homogeneous yet defect-rich metastable nanotwinned structure, which serves as the initial microstructural state for the phase evolution to be studied during subsequent heat treatments.

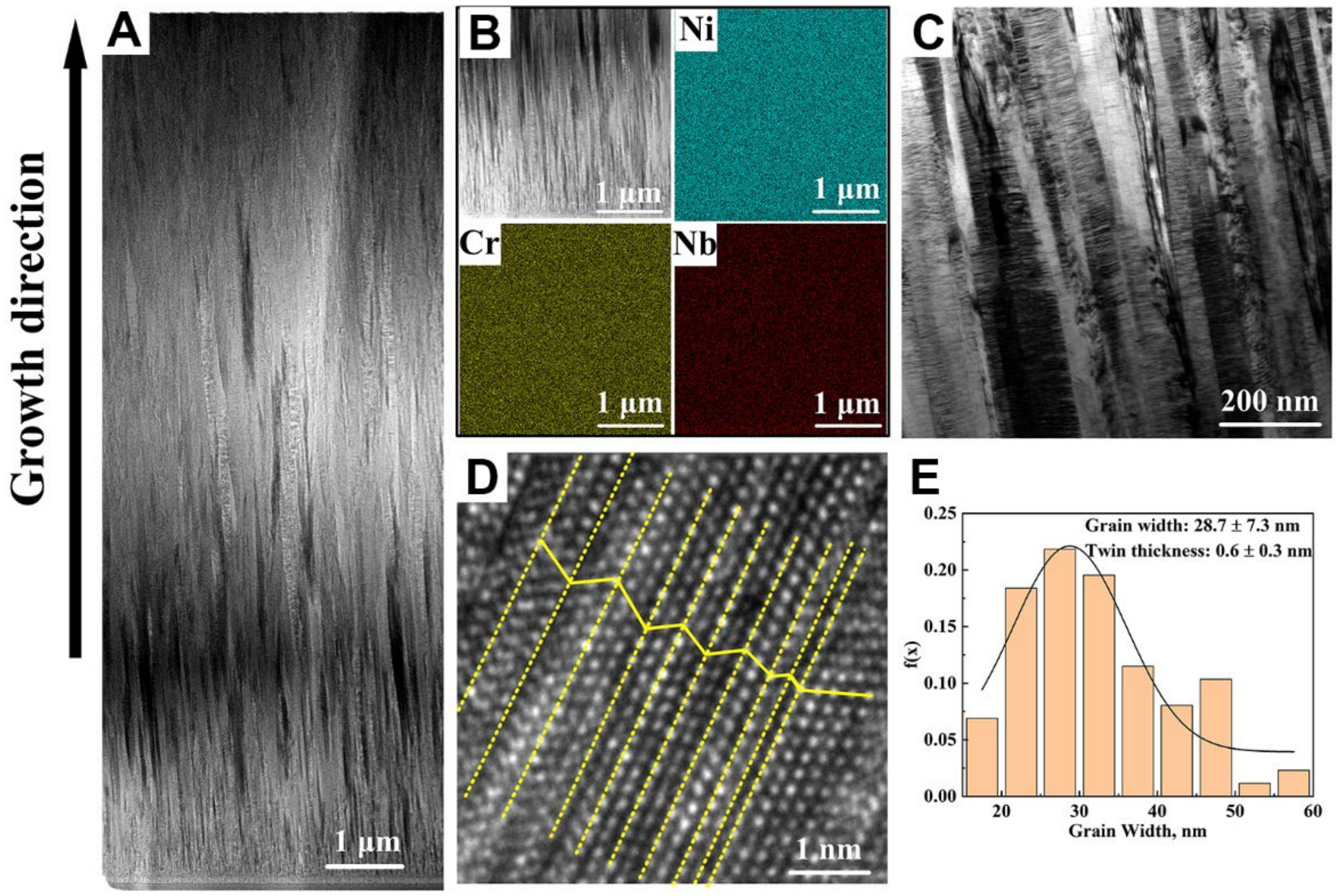

**Fig. 2. (A) BF-STEM image showing the cross-sectional view of the columnar nanotwinned structure of the as-sputtered Inconel 725 sample. (B) BF-STEM image of the bottom region and the corresponding EDS elemental maps of Ni, Cr, and Nb, indicating a uniform elemental distribution. (C) Bright-field TEM**

**image highlighting columnar grains with a high density of nanotwins. (D) High-resolution TEM image showing twin boundaries. (E) Statistical distribution of the columnar grain width and twin thickness.**

## 3.2 Grain growth and precipitation behavior from direct aging treatments

The cross-sectional microstructures of the *Single-aged* and *Dual-aged samples* are first examined to clarify the effect of direct aging without prior solution treatment. While the as-sputtered Inconel 725 film consists of a uniform columnar nanotwinned structure without detectable elemental segregation, direct aging introduces pronounced microstructural heterogeneity along the deposition direction. The corresponding grain sizes and volume fractions of the precipitated phases for both aged conditions are summarized in Table 1.

In the *Single-aged sample* (Fig. 3A), the top region largely retains the columnar nanotwinned structure, with the average columnar width increasing to 161 ± 105 nm (Table 1). In contrast, the middle and bottom regions undergo a pronounced transformation into equiaxed grains, with average grain sizes of 920 ± 361 nm and 243 ± 68 nm, respectively, which is consistent with the nanotwinned-to-abnormally recrystallized microstructural transition reported previously [12]. A qualitatively similar heterogeneous structure is also observed in the *Dual-aged sample* (Fig. 3B), where the top region again retains the nanotwinned morphology with the average columnar width increasing to 123 ± 45 nm, while the middle and bottom regions exhibit abnormal grain growth with average grain sizes of 420 ± 47 nm and 285 ± 24 nm, respectively (Table 1). These observations indicate that direct aging preserves portions of the original nanotwinned architecture while simultaneously driving localized recrystallization and coarsening. This behavior is associated with the residual stress gradient induced during sputtering, with the residual stress being highest near the substrate and progressively decreasing toward the free surface, thereby producing different recrystallization driving forces and nucleation densities across the film thickness [12, 19]. In the top region, the relatively low residual stress provides insufficient driving force for extensive recrystallization, allowing the original columnar nanotwinned structure to be largely retained. In the middle region, the higher driving force activates recrystallization, but the relatively low nucleation density allows selected recrystallized grains to grow abnormally, resulting in pronounced grain coarsening. In contrast, the highest residual stress in the bottom region provides a greater driving force for recrystallization and promotes a higher nucleation density. Competition among the numerous recrystallized grains subsequently limits their growth, leading to the finer equiaxed microstructure observed near the substrate.

Concomitant with this structural evolution, extensive precipitation of Cr-rich precipitates and δ phases occur in both direct aging conditions. In the *Single-aged sample*, Cr-rich precipitates are preferentially distributed along grain boundaries, whereas δ precipitates are observed both at grain boundaries and within the γ grains. Quantitative image analysis shows that the r-rich precipitates and δ-phase fractions reach 10.1% and 8.4% in the top region, 28.5% and 15.5% in the middle region, and 20.4% and 10.3% in the bottom region, respectively (Table 1). Similar precipitation behavior is observed in the *Dual-aged sample*, where the volume percentages of Cr-rich precipitates and δ-phase are 16.5% and 7.8% in the top region, 24.4% and 11.5% in the middle region, and 23.2% and 12.8% in the bottom region, respectively (Table 1). Supplementary diffraction and compositional analyses further confirm the formation of Cr-rich precipitates and Ni- and Nb-rich δ phases (Fig. S1).

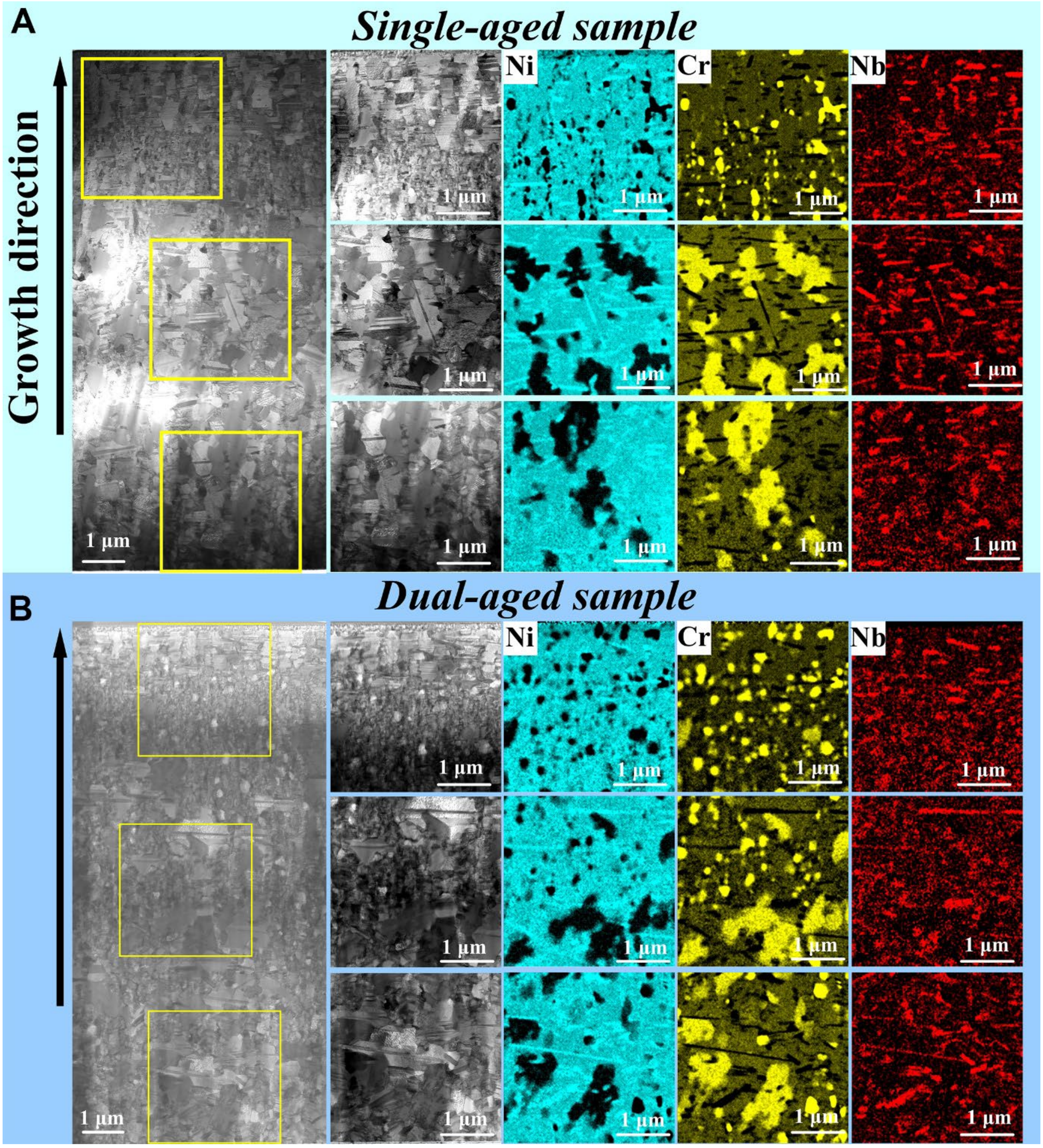

**Fig. 3. BF-STEM images of representative cross-sectional regions of the film subjected to different aging heat treatments, accompanied by EDS maps of Ni, Cr, and Nb highlighting the locations of the Cr-rich precipitates and ($Ni_3Nb$) δ-phase. (A) *Single-aged sample*; (B) *Dual-aged sample*.**

**Table 1 Grain size, volume fractions of precipitated phases, and hardness in sputtered Inconel 725 after different heat treatments. (ND: not detected.)**

| | Region | Grain size, nm | δ phases, % | Cr-rich precipitates, % | γ′/γ″, % | Hardness, GPa |
|---|---|---|---|---|---|---|
| ***Single-aged sample*** | **Top** | **161 ± 105** | **8.4** | **10.1** | **ND** | **8.21 ± 0.17** |
| | **Middle** | **920 ± 361** | **15.4** | **28.5** | | |
| | **Bottom** | **243 ± 68** | **10.3** | **20.4** | | |
| ***Dual-aged sample*** | **Top** | **123 ± 45** | **7.8** | **16.5** | **ND** | **8.38 ± 0.25** |
| | **Middle** | **420 ± 47** | **11.5** | **24.4** | | |
| | **Bottom** | **285 ± 24** | **12.8** | **23.2** | | |
| ***Solutionized sample*** | **Top** | **728 ± 196** | **1.6** | **8.6** | **2.5** | **6.9 ± 0.45** |
| | **Middle** | | **2.6** | **9.1** | | |
| | **Bottom** | | **4.6** | **16.2** | | |
| ***Solutionized + single-aged sample*** | **Top** | **964 ± 251** | **3.9** | **8.7** | **5.4** | **7.54 ± 0.37** |
| | **Middle** | | **4.5** | **8.5** | | |
| | **Bottom** | | **4.9** | **10.8** | | |
| ***Solutionized + dual-aged sample*** | **Top** | **856 ± 249** | **8.3** | **16.8** | **11.7** | **8.92 ± 0.63** |
| | **Middle** | | **7.4** | **14.3** | | |
| | **Bottom** | | **5.7** | **11.7** | | |

In addition to the extensive precipitation of Cr-rich precipitates and δ phases, it is critical to determine whether the primary strengthening γ′/γ″ precipitates form within the γ matrix under direct aging conditions. BF images of representative γ regions (Figs. 4A and 4D), together with corresponding SAED patterns acquired along the high-index $[\bar{1}12]$ zone axis (Figs. 4B and 4E), reveal only face-centered cubic (FCC) reflections from the γ matrix. Consistently, the corresponding DF images (Figs. 4C and 4F) show no evidence of γ′/γ″ precipitation in either the *Single-aged* or *Dual-aged* samples. These results indicate that direct aging alone is insufficient to activate the conventional γ′/γ″ strengthening pathway in sputtered Inconel 725 films, and that under these conditions δ-phase formation is strongly favored over γ′/γ″ precipitation.

In conventional wrought and as-cast Inconel alloys, γ′/γ″ precipitates typically form during relatively low-temperature aging treatments (e.g., 730 °C for 8 h), while δ phase generally forms during higher-temperature aging through γ″ transformation or direct precipitation from the γ matrix (e.g., approximately 100 h at 800 °C or 30 h at 900 °C) [20, 25]. In the sputtered films, however, the high sputtering rate (9.2 nm/s) together with the relatively low stacking fault energy of Inconel alloys (28 – 52 $mJ/m^2$ [26-28]) promotes the formation of a high density of twin boundaries within the columnar grains. These twin boundaries provide crystallographically favorable nucleation sites for δ-phase formation due to the good matching between the {111} twin boundary plane and the δ-phase habit plane [29]. The preferential nucleation of δ phase at twin boundaries is further supported by the observations in Fig. S2.

In addition, deposition-induced residual stresses distort the crystal lattice [30, 31], thereby lowering the nucleation barrier and providing energetically favorable sites for δ-phase precipitation [32]. Similar behavior has been reported in deformation-processed and additively manufactured alloys during low-temperature aging [33, 34]. Consequently, the good crystallographic matching at twin boundaries, together with residual stress induced lattice distortion, lowers the heterogeneous nucleation barrier for δ-phase formation [29]. This promotes preferential δ-phase precipitation and Nb consumption, thereby suppressing the conventional γ′/γ″ precipitation pathway.

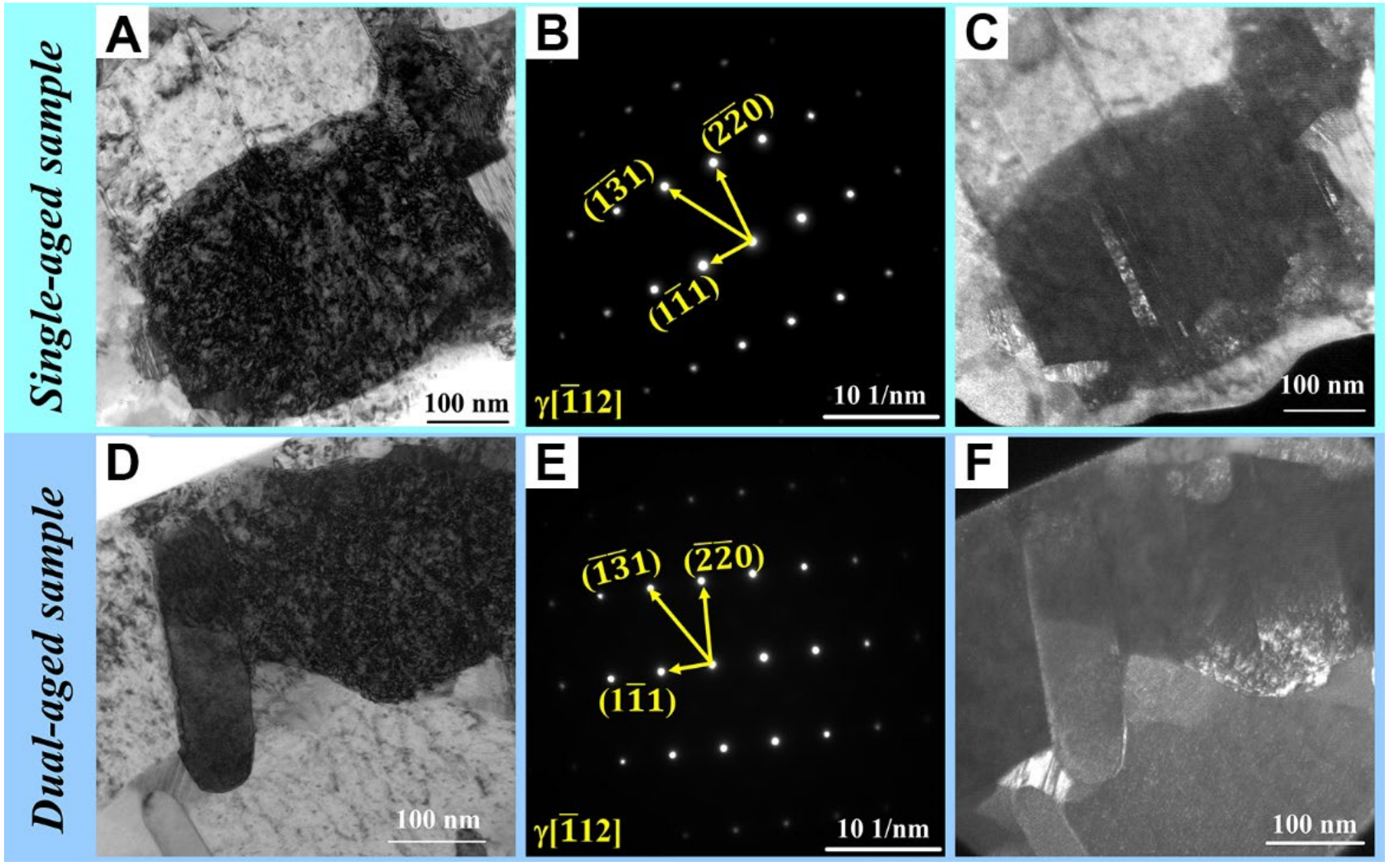

**Fig. 4. (A-C) BF image, corresponding SAED pattern, and DF image of the γ phase in the *Single-aged sample*; (D-F) corresponding images for the *Dual-aged sample*. No γ′/γ″ precipitates are observed in either condition.**

### 3.3 Grain growth and precipitation from Solutionizing and Solutionizing + Aging treatments

Given that direct aging treatment alone is insufficient to activate $\gamma'/\gamma''$ precipitation due to the preferential formation of δ-phase and Cr-rich precipitates, a high-temperature solution treatment (1038 °C for 2 h) was employed to modify the subsequent precipitation pathway. The corresponding grain sizes and volume fractions of the precipitated phases for these three heat-treatment conditions are summarized in Table 1.

#### 3.3.1 Microstructure of the *Solutionized* sample

The cross-sectional BF-STEM image of the *Solutionized sample* (Fig. 5A) reveals a uniform equiaxed microstructure throughout the film thickness, with an average grain size of 728 ± 196 nm (Table 1), indicating that the high-temperature solution treatment eliminates the pronounced through-thickness microstructural heterogeneity observed after direct aging. The original nanotwinned structure is largely removed during recrystallization. Notably, despite the significantly higher treatment temperature, the resulting grain size remains smaller than the abnormally recrystallized grains observed in the middle region of the *Single-aged sample*. This suggests that the initially high density of twin boundaries suppresses excessive grain growth during high-temperature exposure.

The corresponding EDS elemental maps of Ni, Cr, and Nb show that both Cr-rich precipitates and δ phases remain present after solution treatment, although their distribution and volume fractions differ significantly from those observed under direct aging conditions. Specifically, Cr-rich precipitates are primarily distributed along γ grain boundaries, while Ni- and Nb-enriched δ-phase exhibits a characteristic lamellar morphology within the γ grains. Quantitative image analysis shows that the Cr-rich precipitates volume fractions are reduced to 8.6%, 9.1%, and 16.2% in the top, middle, and bottom regions, respectively, while the corresponding δ-phase volume fractions are significantly reduced to 1.6%, 2.6%, and 4.6% compared with the directly aged samples (see Table 1). The substantial reduction in Cr-rich precipitates and especially δ-phase precipitation indicates that fewer alloying elements are consumed by these competing phases during solution treatment, increasing solute availability for subsequent precipitation within the γ matrix.

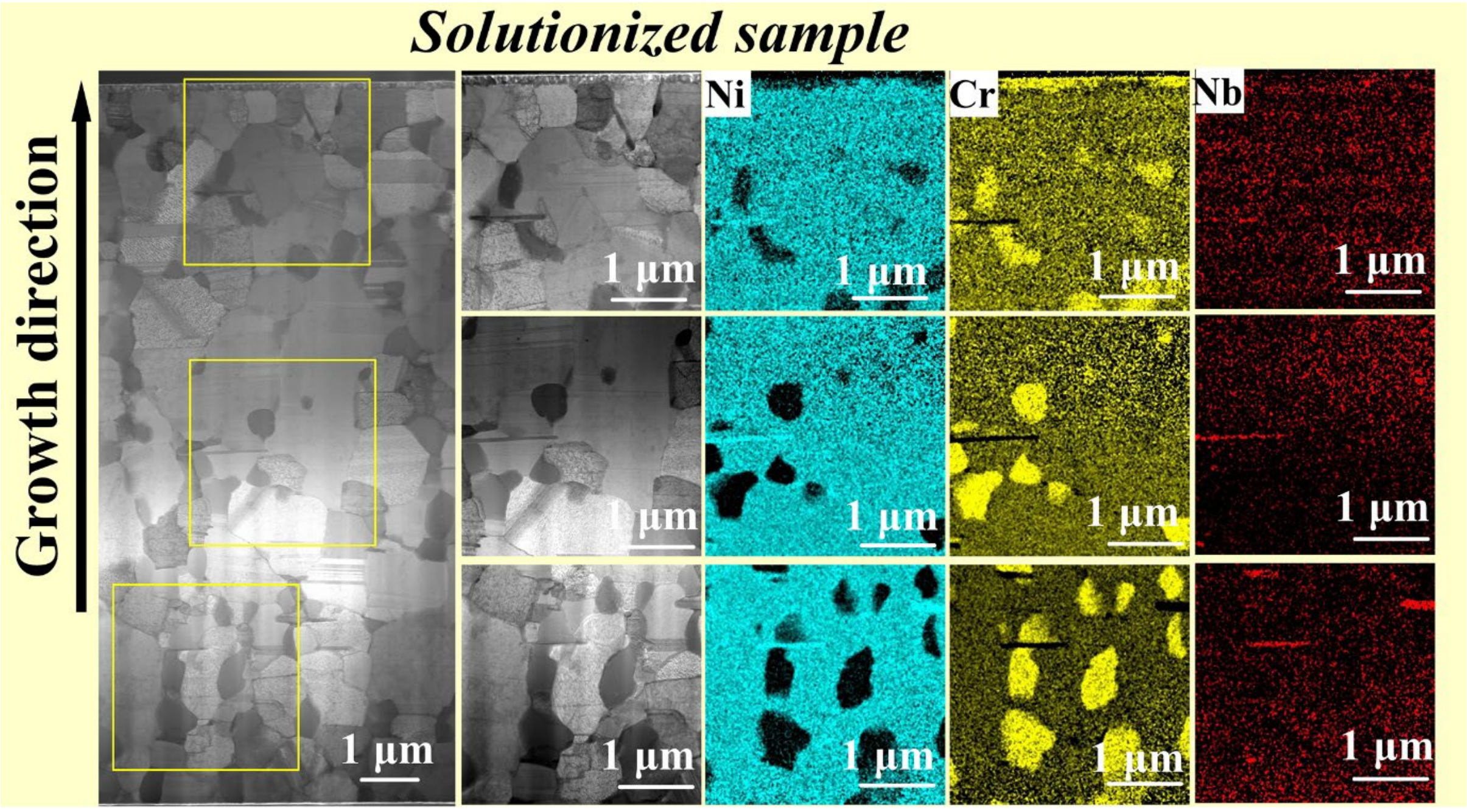

**Fig. 5. BF-STEM images of representative cross-sections showing the top, middle, and bottom layers through the film thickness of the *Solutionized sample*, illustrating the microstructural evolution after solution treatment. EDS maps of Ni, Cr, and Nb indicate the distribution of the Cr-rich precipitates and δ-phase.**

To further confirm the formation of γ′/γ″ precipitates, detailed TEM analysis was performed using BF imaging, SAED under multiple diffraction conditions, and CDF imaging, as presented in Fig. 6. Fig. 6A shows a BF TEM image of the *Solutionized sample*, in which a rafted contrast is observed without lamellar δ phases inside the γ grains. SAED patterns acquired from the γ matrix along the low-index [001] and [011] zone axes (Figs. 6B and 6C) exhibit only FCC reflections corresponding to the γ phase. However, under the high-index $[\bar{1}12]$ zone axis (Fig. 6D), additional superlattice reflections associated with the γ′/γ″ phases become visible, as indicated by the green arrows, confirming their precipitation within the γ matrix. A corresponding CDF image (Fig. 6E) taken using the $[\bar{1}12]$ superlattice reflections reveals a high density of ultrafine, spherical γ′/γ″ precipitates uniformly embedded in the γ matrix. Statistical analysis shows that these precipitates have an average size of approximately 3 nm (Fig. 6F), indicating an early-stage and very fine-grained precipitation state.

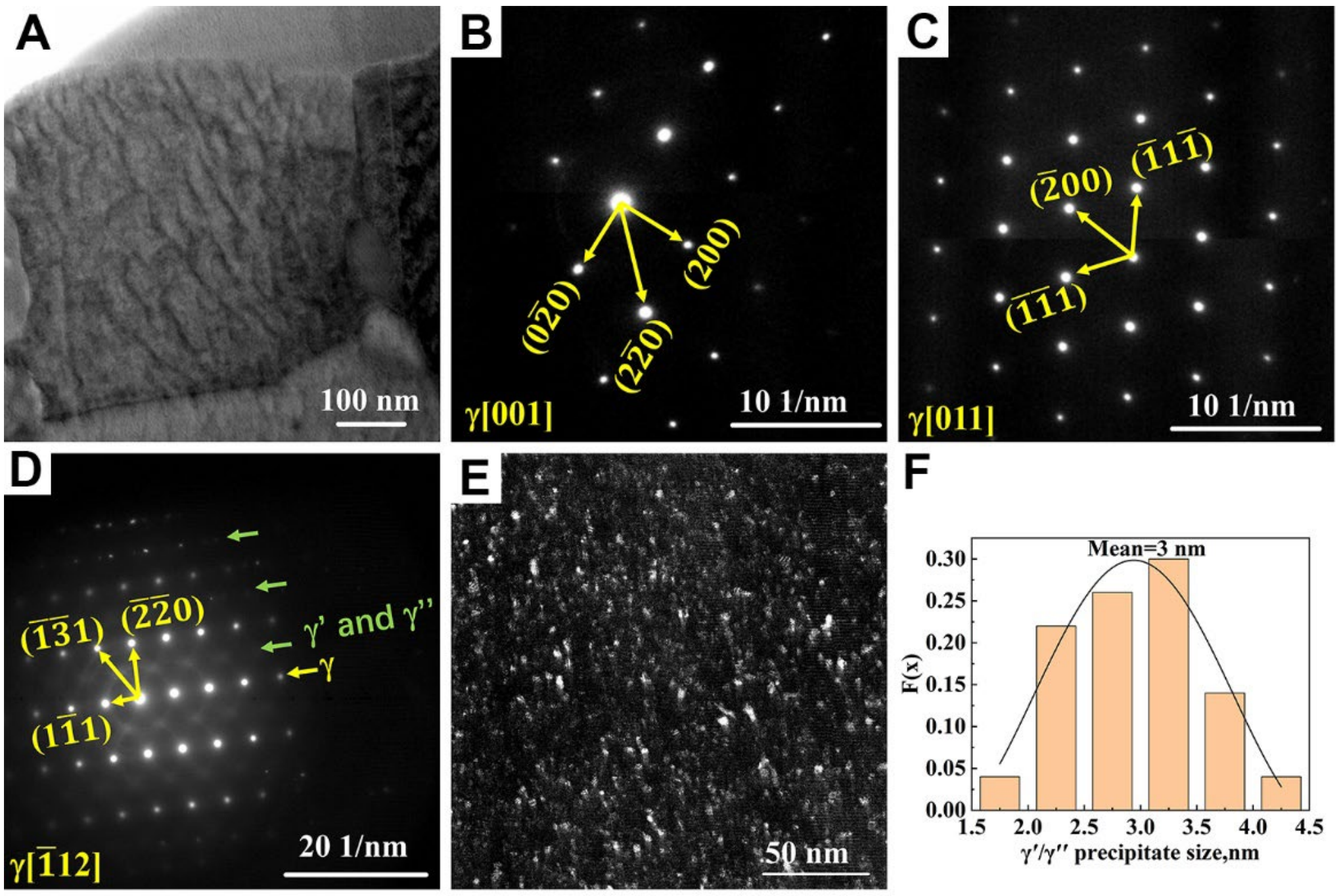

**Fig. 6.** **(A) BF image of the γ phase in the *Solutionized sample*. (B–D) SAED patterns taken along the [001], [011], and [$\bar{1}$12] zone axes within the γ matrix. Only the [$\bar{1}$12] orientation shows the γ′/γ″ diffraction spots within the γ matrix. (E) Centered dark-field (CDF) image of the [$\bar{1}$12] zone axes of the γ phase showing γ′/γ″ precipitates. (F) Size distribution of γ′/γ″ precipitates obtained from CDF image analysis.**

BF-STEM imaging and EDS elemental maps further characterize the γ′/γ″ precipitation behavior in the *Solutionized sample*, as presented in Fig. 7. Fig. 7A reveals a homogeneous distribution of γ′/γ″ precipitates within the γ matrix. However, the corresponding EDS maps of Ni, Cr, Mo, Fe, Nb, Ti, and Al (Figs. 7B–H, respectively) display a uniform elemental distribution across the γ matrix, indicating that the precipitates remain too fine to produce measurable compositional segregation at this stage.

The role of this “solutionizing” heat treatment in the present system differs from its conventional purpose. In wrought and cast Inconel alloys, solutionizing is typically used to homogenize elemental distribution and dissolve pre-existing phases [35]. This step may even be completely skipped in additively manufactured alloys due to their relatively uniform elemental distribution and reduced segregation compared to their cast and wrought counterparts [20]. However, as shown earlier in Fig. 2, the as-sputtered films in this study already exhibit a uniform elemental distribution with no detectable segregation. Instead, the primary role of the solutionizing heat treatment here is to reconstruct the microstructure through recrystallization and grain growth, effectively eliminating the columnar nanotwinned structure and providing deposition-induced residual stress relief. Similar behavior has been reported in several additively manufactured Inconel alloys, where high-temperature solution treatment modifies their columnar grain structures into fully equiaxed morphologies and removes residual stresses [34, 36, 37]. As a result, solution treatment alters the precipitation pathway and enables $\gamma'/\gamma''$ formation. Notably, despite the elimination of nanotwins during solution treatment, the resulting $\gamma$ matrix retains an ultrafine-grained structure and thus a relatively high grain boundary density compared with conventionally solution-treated alloys [9, 38, 39].

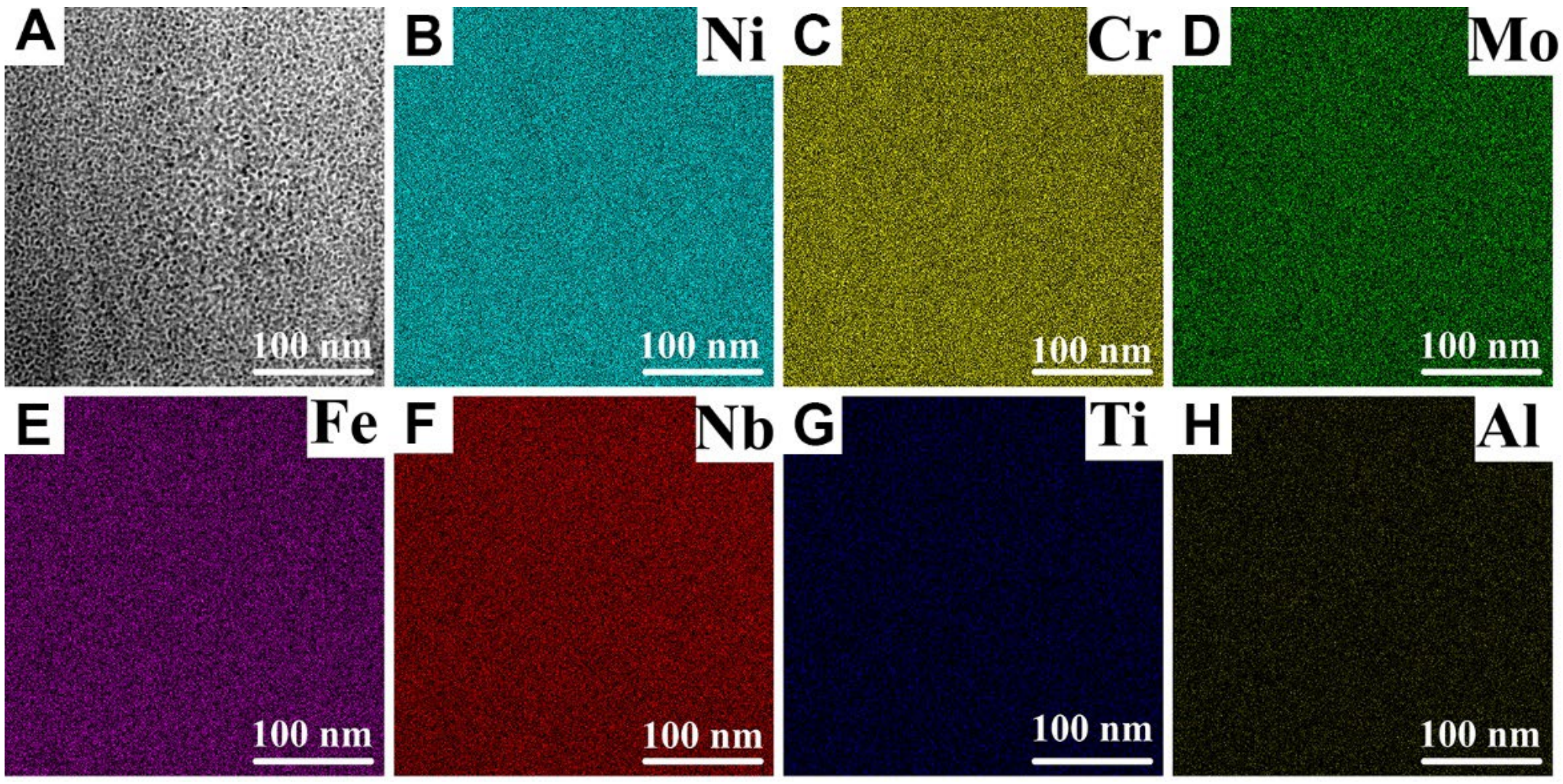

**Fig. 7. (A) BF-STEM image of the γ-phase region in the *Solutionized sample*. (B-H) EDS maps of Ni, Cr, Mo, Fe, Nb, Ti, and Al showing a uniform elemental distribution.**

### 3.3.2 Microstructure of the *Solutionized + single-aged* sample and *Solutionized + dual-aged sample*

Aging treatments applied subsequently to the solutionization step further drive the evolution of the precipitate phases in a manner that differs markedly from the direct aging response. In the *Solutionized + single-aged sample* (Fig. 8A), the microstructure retains a relatively uniform equiaxed grain structure with an average grain size of 964 ± 251 nm. EDS elemental maps of Ni, Cr, and Nb reveal that δ-phase precipitation becomes more pronounced compared with the *Solutionized sample*, while Cr-rich precipitates fractions remain relatively stable. The Cr-rich precipitates volume fractions remain nearly constant across the film thickness (8.7% in the top region, 8.5% in the middle region, and 10.8% in the bottom region), whereas the δ-phase volume fractions increase to 3.9%, 4.5%, and 4.9%, respectively (see

Table 1). The slightly higher volume fraction of Cr-rich precipitates in the bottom region is consistent with the through-thickness variation already present in the *Solutionized sample*, suggesting that this variation may be inherited from the solution-treated microstructure rather than newly introduced during the subsequent single-aging treatment. Notably, δ precipitates undergo a clear morphological evolution during this aging treatment. Although lamellar δ precipitates within γ grains are still present, their fraction decreases and δ precipitates become increasingly concentrated along γ grain boundaries.

Compared with the average grain size of 728 ± 196 nm in the *Solutionized* sample, only modest grain coarsening occurs during subsequent aging at 730 °C. The high grain boundary area and associated interfacial energy of the ultrafine-grained solutionized microstructure provide a thermodynamic driving force for grain growth. However, the concurrent presence of δ phase and Cr-rich precipitates may restrict grain boundary migration through a pinning effect, thereby preventing extensive grain growth. Therefore, the observed modest coarsening likely reflects a balance between the thermodynamic driving force for grain growth and precipitate-induced grain boundary pinning. The redistribution of δ precipitates from grain interiors toward grain boundaries may also reduce direct competition between δ and γ′/γ″ for intragranular nucleation sites within the γ matrix.

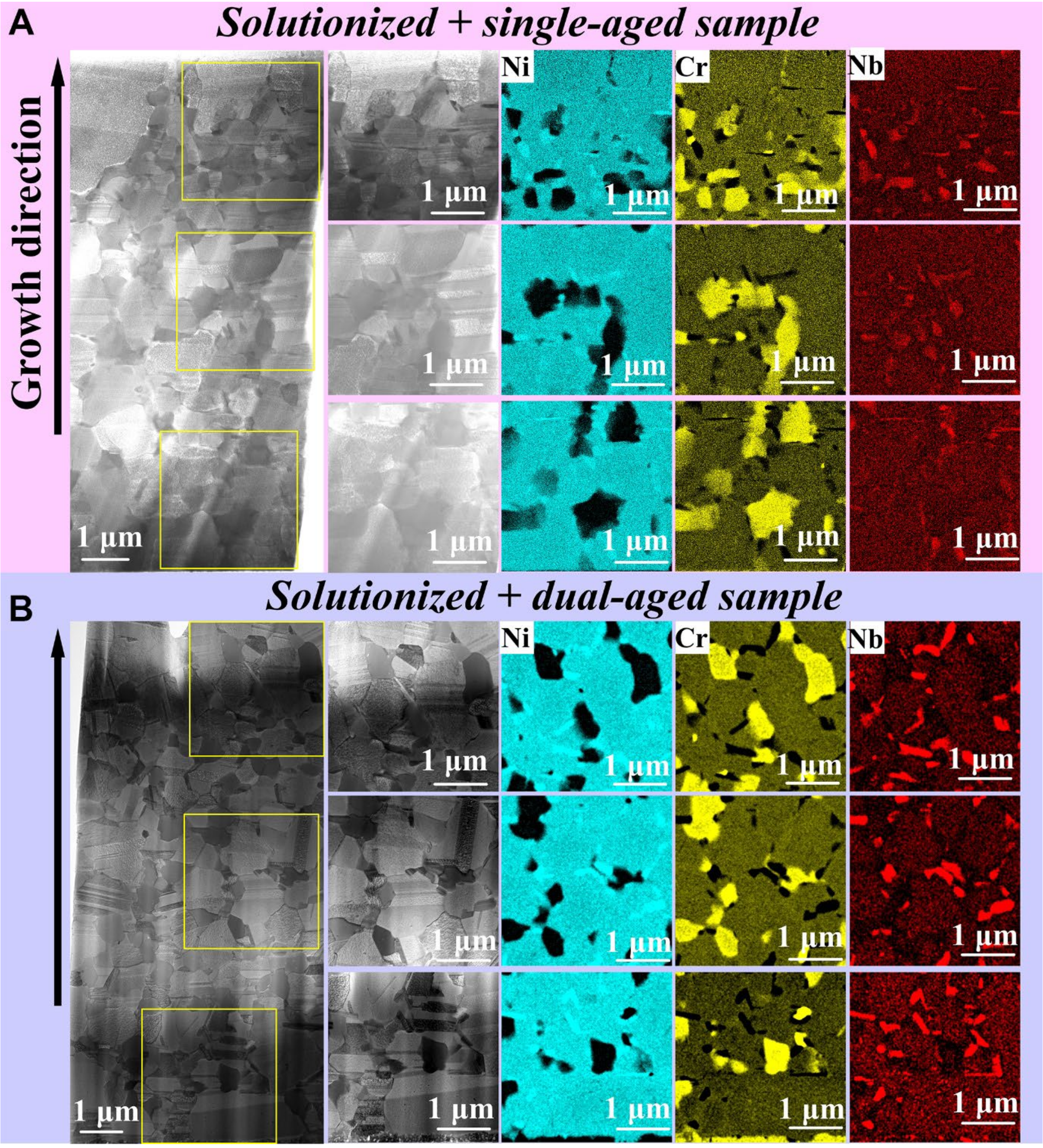

**Fig. 8.** Representative BF-STEM cross-sectional images showing the top, middle, and bottom regions throughout the film thickness of the (A) *Solutionized* + *Single-aged* and (B) *Solutionized* + *Dual-aged* samples, illustrating the microstructural evolution. Corresponding EDS maps of Ni, Cr, and Nb reveal the distributions of Cr-rich precipitates and the δ phase.

Dual aging after solutionization (Fig. 8B) leads to continued evolution of δ precipitation and elemental partitioning. The microstructure remains equiaxed with an average grain size of 856 ± 249 nm. Compared with the *Solutionized* and *Solutionized + single-aged* samples, EDS maps show significantly stronger Cr-rich precipitates and δ-phase segregation, suggesting that dual aging enhances elemental partitioning. The Cr-rich precipitates volume fractions (16.8%, 14.3%, and 11.7%) remain comparable to those of the *Solutionized sample*, while δ-phase fractions further increase to 8.3%, 7.4%, and 5.7% across the film thickness (see Table 1). More importantly, δ precipitation becomes increasingly localized along γ grain boundaries rather than remaining within the γ grains. The average short-axis size of the grain boundary δ phases was 148 ± 52 nm, and many of these precipitates exhibit elongated morphologies along the grain boundaries. This progressive redistribution from intragranular lamellar precipitates to grain boundary-dominated precipitation indicates that grain boundaries become the primary nucleation sites for δ phases during prolonged aging.

To further understand the precipitation behavior within the γ matrix, the evolution of γ′/γ″ precipitates is examined for both the *Solutionized + single-aged* and *Solutionized + dual-aged* samples (Fig. 9). In the *Solutionized + single-aged sample* (Figs. 9A1–D1), γ′/γ″ precipitates are identified through diffraction analysis. Under the low-index [001] zone axis (Fig. 9B1), only FCC reflections corresponding to the γ matrix are observed, indicating that γ′/γ″ reflections remain weak under this condition. In the *Solutionized + dual-aged sample*, however, distinct $\gamma'/\gamma''$ superlattice reflections become clearly visible under the same [001] orientation (Fig. 9B2), indicating further development of these precipitates. Under the [011] zone axis (Fig. 9C1), additional diffraction spots associated with $\gamma'/\gamma''$ phases appear in the *Solutionized + single-aged sample*, while similar reflections are also observed in the *Solutionized + dual-aged sample* (Fig. 9C2). More distinct $\gamma'/\gamma''$ superlattice reflections are observed under the high-index $[\bar{1}12]$ zone axis in the *Solutionized + single-aged sample* (Fig. 9D1), whereas in the *Solutionized + dual-aged sample*, the corresponding high-index [013] orientation provides even clearer diffraction features (Fig. 9D2). Corresponding CDF imaging further reveals the morphological evolution of $\gamma'/\gamma''$ precipitates. In the *Solutionized + single-aged sample* (Fig. 9E1), $\gamma'/\gamma''$ precipitates transform from the ultrafine spherical particles observed in the *Solutionized* condition into well-defined lenticular morphologies, with average dimensions of ~11.5 nm and ~22.7 nm (Fig. 9F1) in the minor and major axes. In the *Solutionized + dual-aged sample* (Fig. 9E2), lenticular precipitates also dominate and have similar dimensions (~9.5 nm and ~22.5 nm), but exhibit a more uniform distribution within the $\gamma$ matrix and local enrichment near grain boundaries. These observations indicate that subsequent aging transforms the $\gamma'/\gamma''$ precipitates from ultrafine spherical particles into a lenticular morphology, while dual aging further promotes a more uniform distribution of the lenticular precipitates within the $\gamma$ matrix.

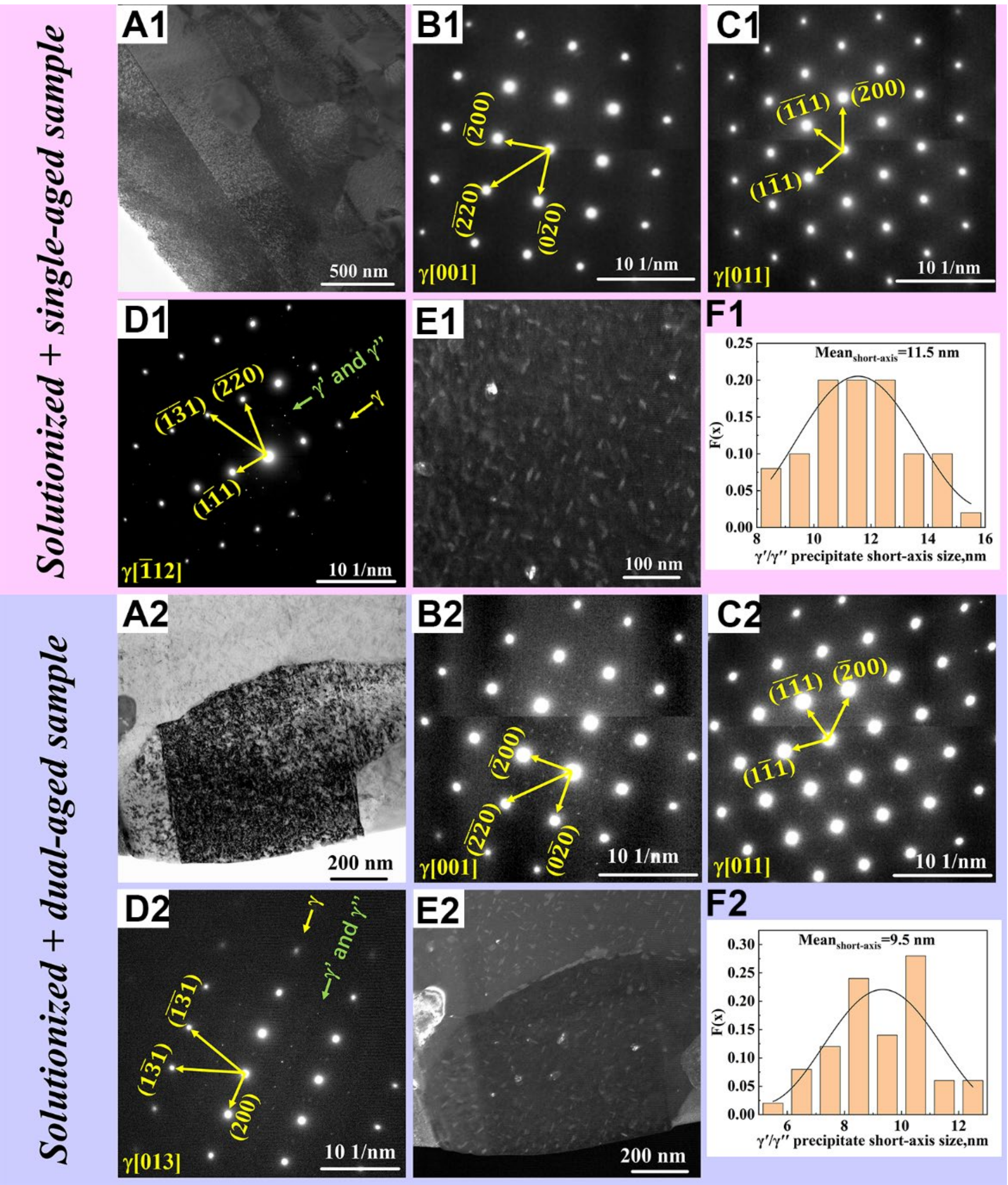

**Fig. 9.** BF images, SAED patterns, and CDF images showing the evolution of γ′/γ″ precipitates in the *Solutionized + single-aged* and *Solutionized + dual-aged samples*. (A1–F1) Results for the *Solutionized + single-aged sample*: (A1) BF image of the γ phase; (B1–D1) SAED patterns along the [001], [011], and [$\bar{1}$12] zone axes, with γ′/γ″ superlattice reflections visible in the [011] and [$\bar{1}$12] orientations; (E1) CDF image; (F1) short-axis size distribution of γ′/γ″ precipitates. (A2–F2) Corresponding results for the *Solutionized + dual-aged sample*: (A2) BF image; (B2–D2) SAED patterns along the [001], [011], and [013] zone axes, all **showing clear $\gamma'/\gamma''$ superlattice reflections; (E2) CDF image; (F2) short-axis size distribution of $\gamma'/\gamma''$ precipitates.**

This structural evolution is accompanied by progressive changes in chemical partitioning behavior, as further illustrated in Fig. 10. BF-STEM images (Figs. 10A1 and 9A2) reveal a clear evolution in precipitate morphology. In the *Solutionized + single-aged sample* (Fig. 10A1), lenticular $\gamma'/\gamma''$ precipitates are non-uniformly distributed within the $\gamma$ matrix, with locally enhanced precipitation in certain regions, suggesting that this condition represents a transitional stage. With further aging, the *Solutionized + dual-aged sample* (Fig. 10A2) exhibits more well-developed lenticular $\gamma'/\gamma''$ precipitates with two distinct orientations, indicating a more advanced stage of precipitate evolution. Corresponding EDS elemental maps (Figs. 10B1–H1 and 10B2–H2) further reveal the evolution of compositional characteristics. In the *Solutionized + single-aged* sample, compositional variations are observed relative to the *Solutionized* sample, with the Cr content decreasing from 24.0 to 17.5 wt.% and the Nb content increasing from 3.3 to 6.7 wt.%. These changes indicate the onset of elemental partitioning associated with $\gamma'/\gamma''$ precipitation. With continued aging, the *Solutionized + dual-aged sample* shows significantly enhanced compositional segregation, particularly Nb and Ti enrichment, accompanied by Cr and Fe depletion. In particular, the Cr content decreases from 24 wt.% to 13.2 wt.%, whereas the Nb content increases from 3.3 wt.% to 10.7 wt.%. These observations indicate that $\gamma'/\gamma''$ precipitates evolve through coupled structural and chemical changes during aging, transitioning from a non-uniform distribution with weak compositional fluctuations to a more developed state characterized by well-defined morphology and pronounced elemental partitioning within the γ matrix.

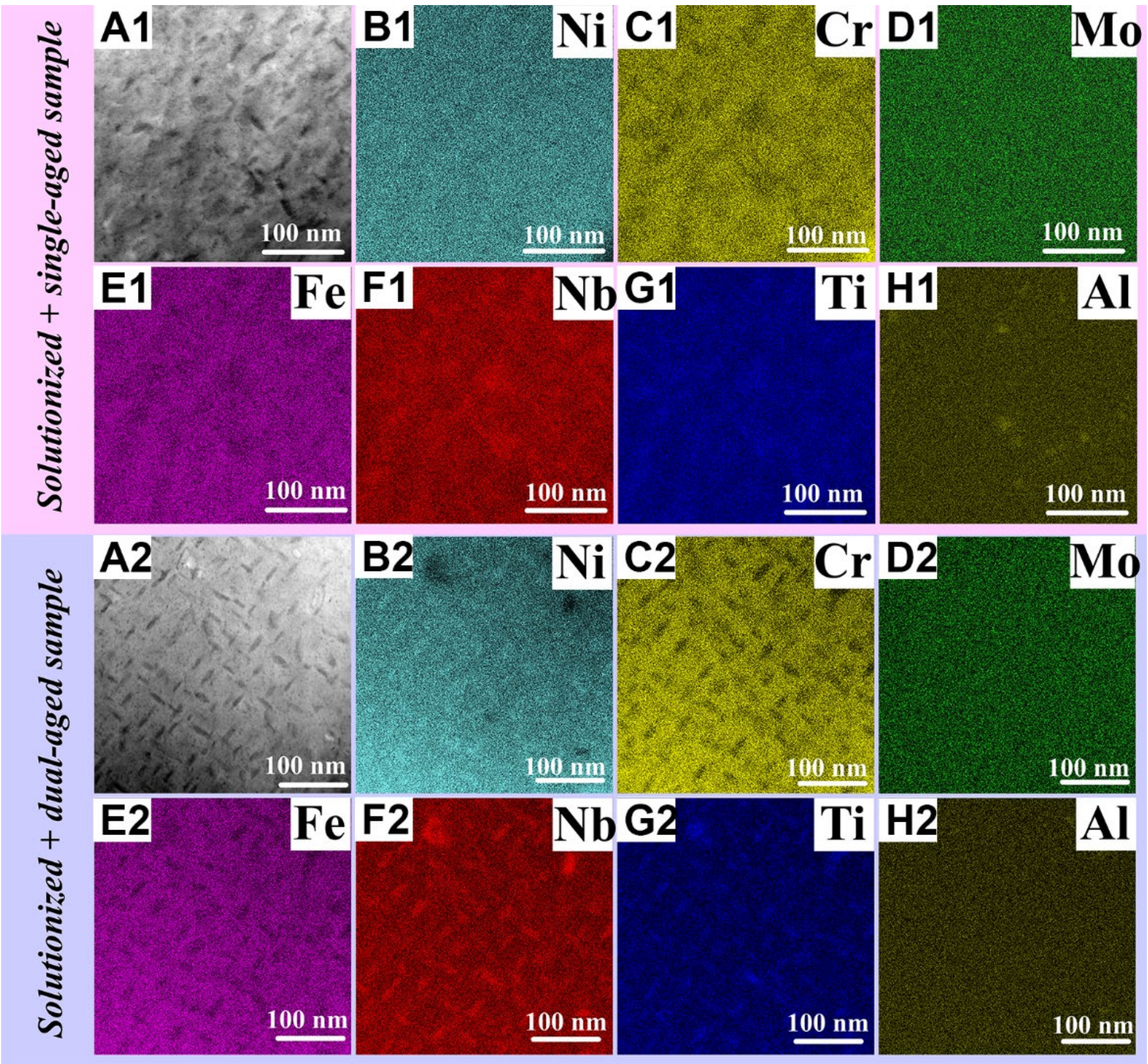

**Fig. 10.** **(A1–H1) BF-STEM image and corresponding EDS maps (Ni, Cr, Mo, Fe, Nb, Ti, and Al) of the γ-phase region in the *Solutionized + single-aged sample*, showing slight elemental segregation. (A2–H2) Corresponding results for the *Solutionized + dual-aged sample*, revealing significantly enhanced elemental segregation associated with γ′/γ″ precipitates.**

Solution treatment induces recrystallization in the sputtered Inconel 725, leading to the formation of ultrafine equiaxed grains and the elimination of twin boundaries, which significantly reduces the δ phase fraction and promotes the formation of γ′/γ″ precipitates, thereby altering the phase evolution during subsequent aging. During aging, the high density of grain boundaries in the refined γ matrix promotes preferential nucleation of δ phases along grain boundaries, which are known to provide a higher driving force for nucleation [40-42]. These grain boundaries δ phases stabilize the microstructure through boundary pinning, thereby inhibiting grain growth and maintaining the refined γ grain size. Notably, the intragranular δ phase formed during solution treatment gradually diminishes during aging, as the removal of intragranular δ phases releases solute and eliminates competing nucleation sites within the γ matrix. This process promotes the morphological evolution of γ′/γ″ precipitates from spherical to lenticular, accompanied by an increase in their volume fraction. This behavior demonstrates that the dissolution of intragranular δ phases directly facilitates the evolution and strengthening effect of γ′/γ″ precipitates. This concurrent evolution highlights a spatial redistribution of precipitation sites, which governs the partitioning of γ′/γ″ and δ phases between intragranular regions and grain boundaries.

These results suggest a transition in phase segregation behavior from defect-dominated to spatially partitioned precipitation, in which the redistribution of interfaces governs phase selection. As a result, the conventional competitive relationship between γ′/γ″ and δ phases is effectively decoupled, enabling the simultaneous development of intragranular strengthening and grain boundary stabilization.

### 3.4 Evolution of hardness with heat treatment

Hardness is an important indicator of the mechanical response to microstructural evolution. Owing to its high density of twin boundaries, sputtered Inconel 725 alloy exhibits distinctive grain growth and precipitation behaviors during heat treatment compared with conventionally processed Inconel 725 alloy. Consequently, it is important to correlate the hardness evolution under different heat-treatment routes with the corresponding microstructural evolution.

In the present study, the as-sputtered Inconel 725 film exhibits a columnar nanotwinned architecture (>95% nanotwinned grains) with a supersaturated $\gamma$ matrix. Owing to the extremely small columnar width (~28.7 nm) and twin thickness (~0.6 nm), this microstructure delivers an exceptionally high hardness of ~10.58 GPa (Fig. 10). In contrast, conventionally processed coarse-grained Inconel 725 alloys with similar chemical compositions generally exhibit substantially lower hardness close to 4 GPa [14, 43]. According to the solid solution strengthening analysis presented in Supplementary Table S1, the estimated hardness contribution from solid solution strengthening is less than 1 GPa. Therefore, the substantially higher hardness of the as-sputtered film is primarily attributed to its unique nanostructure, particularly the high density of twin boundaries and nanoscale columnar grains.

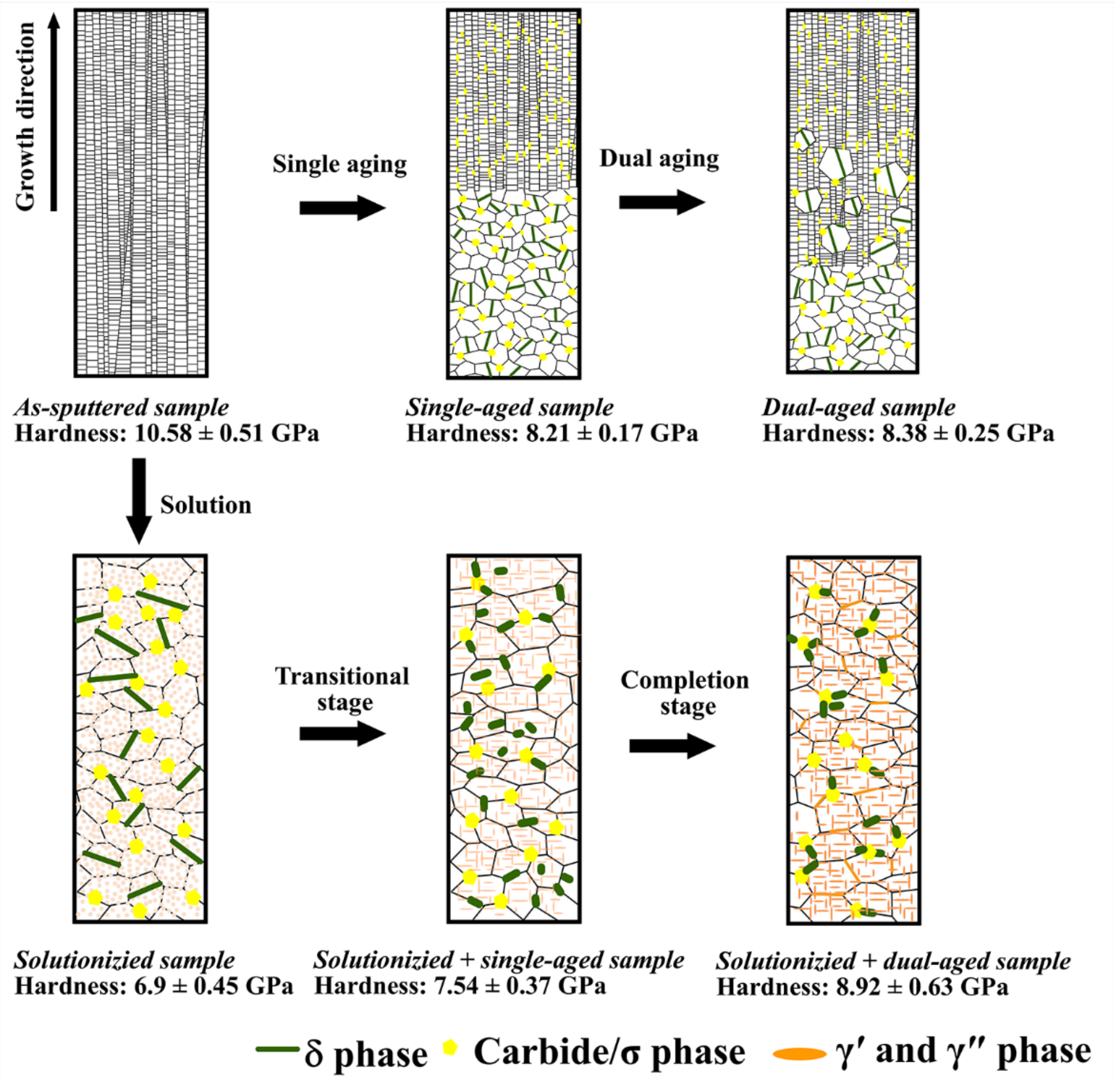

**Fig. 11. Schematic overview of the microstructural evolution of sputtered Inconel 725 under different heat-treatment conditions, illustrating changes in grain morphology, precipitate formation, and hardness. The arrow indicates the deposition direction. Direct aging results in Cr-rich precipitates and δ-phase precipitation while partially retaining the nanotwinned structure and high hardness. Solution treatment enables γ′/γ″ precipitation, and subsequent aging promotes their growth, elemental segregation, and a more uniform distribution, leading to increased hardness.**

Under the direct aging conditions, the dense ultrafine nanotwins are largely preserved due to the significantly lower energy of coherent twin boundaries compared with high-angle grain boundaries, which suppresses boundary migration during aging [44-46]. At the same time, numerous δ-phase precipitates form along the twin boundaries (Fig. S2), providing a pinning effect, consistent with previous reports in Mg alloys and Fe-based superalloys containing dense ultrafine twins [46, 47]. As a result, both single- and dual-aging largely maintain the original nanotwinned framework in the top region (~4 μm, ~30% of the film thickness). This retention is attributed to the high thermal stability of nanotwins and the relatively low residual stress and stored energy near the free surface, which provide insufficient driving force for extensive detwinning and recrystallization during aging [12, 19]. The formation of extensive δ phase depletes Nb from the γ matrix, thereby reducing the supersaturation required for γ′/γ″ precipitation [48], which suppresses γ′/γ″ precipitation under direct aging conditions. Nevertheless, the retained high density of twin boundaries and refined columnar grains sustains relatively high hardness values of 8.21 and 8.38 GPa for the *Single-aged* and *Dual-aged* samples, respectively (Fig. 11).

The nanotwinned structure becomes unstable at the solution-treatment temperature and is largely eliminated by recrystallization, resulting in equiaxed γ grains with a refined size (< 1 μm). The microstructural reconstruction after solution treatment significantly decreases the volume fraction of intragranular δ phases (2.9%, Fig. 5), thereby activating the precipitation of intragranular, spherical γ′/γ″ phases with a volume fraction of 2.5% (Fig. 6). Compared with the columnar nanotwinned structure, where γ′/γ″ precipitation is suppressed, grain coarsening and the low volume fraction of γ′/γ″ during solution treatment reduces hardness to 6.9 GPa (Fig. 11), although this value is still higher than that of conventional wrought and as-cast alloys [14, 49-51].

Subsequent single- and dual-aging treatments further modify the precipitation behavior. Intragranular δ phases progressively disappear, while δ phases and carbides increasingly segregate along grain boundaries with higher volume fractions, providing a pinning effect that stabilizes the refined grain structure. Meanwhile, the removal of intragranular δ phases promotes the growth of γ′/γ″ precipitates within the γ matrix, driving their morphological evolution from spherical to lenticular and increasing their volume fraction to 5.4%, representing a 116% increase compared with the *Solutionized* sample (Figs. 9–10). With dual aging, γ′/γ″ precipitates become well-defined and more uniformly distributed, accompanied by pronounced elemental partitioning and a higher volume fraction of 11.7%, corresponding to a 368% increase relative to the *Solutionized* sample and a 117% increase relative to the *Solutionized + single-aged* sample (Figs. 9–10). Although the grain sizes and estimated solid-solution strengthening contributions of the γ matrix remain relatively similar among the *Solutionized*, *Solutionized + single-aged*, and *Solutionized + dual-aged* samples, the increased volume fraction and development of γ′/γ″ precipitates enhance precipitation strengthening after aging. Consequently, the hardness increases to 7.54 GPa in the *Solutionized + single-aged sample* and reaches a maximum of 8.92 GPa in the *Solutionized + dual-aged sample* (Fig. 11). The higher hardness of the *Solutionized + dual-aged sample* compared with the direct aged samples demonstrates that the synergistic combination of precipitation strengthening and grain refinement, enabled by the spatial redistribution of precipitates, provides more effective strengthening than the columnar nanotwinned structure.

**4. Conclusions**

In this work, the microstructure, phase evolution and mechanical response of sputtered nanotwinned Inconel 725 films under different heat-treatment pathways were systematically investigated. The main conclusions are as follows:

(1) Direct aging of the as-sputtered nanotwinned structure produces a heterogeneous microstructure with retained nanotwinned columnar grains in the top region. The top-region columnar grain width increases to 161 nm after single aging and 123 nm after dual aging. Extensive δ-phase precipitation at twin boundaries and grain boundaries suppresses the formation of the strengthening γ′/γ″ phases through Nb depletion.

(2) High-temperature solution treatment transforms the heterogeneous microstructure into a uniform ultrafine-grained equiaxed microstructure with an average grain size of 728 nm, fundamentally changing the precipitation pathway from δ-phase-dominated precipitation to γ′/γ″ precipitation within the γ matrix.

(3) Subsequent aging promotes the development of uniformly distributed γ′/γ″ precipitates within the γ matrix while retaining δ phases at grain boundaries and maintaining a uniform ultrafine-grained equiaxed microstructure, with average grain sizes of 964 and 856 nm in the *Solutionized + single-aged* and *Solutionized + dual-aged* samples, respectively, thereby providing a favorable microstructural configuration for strengthening.

(4) The hardness evolution reflects a transition in dominant strengthening mechanisms from nanotwin-mediated strengthening to precipitation-controlled strengthening. The highest hardness (~8.9 GPa) among the heat treated samples is achieved after dual aging following solution treatment, where optimized $\gamma'/\gamma''$ precipitation and grain boundary $\delta$-phase pinning act synergistically.

## Acknowledgments

This research was supported by the National Science Foundation (NSF) within the Designing Materials to Revolutionize and Engineer our Future (DMREF) program under Grant Numbers: 2323936 (AMH) and 2502667 (TJR).

**Supplementary Materials**

As shown in Figs. 3 and 5, two types of secondary precipitates are observed in all five heat-treated conditions. To identify these precipitates, selected-area electron diffraction (SAED) and STEM-EDS analyses were performed.

As shown in Fig. S1A and B, one type of precipitate is enriched in Ni, Nb, and Ti. Compared with the surrounding γ matrix, the Ni content increases from 52.6 to 71.4 wt.%, the Nb content from 3.3 to 17 wt.%, and the Ti content from 1.1 to 4.8 wt.%. Based on this compositional enrichment and the corresponding diffraction analysis, these precipitates are identified as the δ phase. This identification is further supported by the SAED pattern acquired along the [110] zone axis of the γ matrix (Fig. S1C), where the characteristic [100] superlattice reflections associated with the δ phase are clearly observed, consistent with previous reports [1, 2].

The second type of precipitate is enriched in Cr, with some regions also exhibiting Mo enrichment (Fig. S1B). In the Cr-rich regions, the Cr content increases from approximately 24 wt.% in the initially homogeneous γ matrix of the as-sputtered sample to approximately 80 wt.%. In the Cr–Mo-rich regions, the Cr content is approximately 36 wt.%, while the Mo content increases from approximately 8.1 wt.% in the as-sputtered γ matrix to 23.6 wt.%. These Cr-rich precipitates exhibit diffraction features that are comparable to those reported for both $M_{23}C_6$ carbides (FCC, face-centered cubic) and σ (TCP, topological close-packed) phase (Fig. S1D and E). However, STEM-EDS mapping does not reveal obvious carbon enrichment in these precipitates, while $M_{23}C_6$ carbides can also exhibit Mo enrichment [3]. Therefore, the EDS results alone do not permit an unambiguous phase identification. Since both $M_{23}C_6$ carbides and σ phase are Cr-rich brittle phases commonly reported in Inconel 725 alloys, and the primary objective of this work is to elucidate the competition between δ-phase precipitation and γ′/γ″ strengthening during different heat treatments rather than to investigate the crystallography of Cr-rich precipitates, these precipitates are conservatively referred to as Cr-rich precipitates throughout the manuscript.

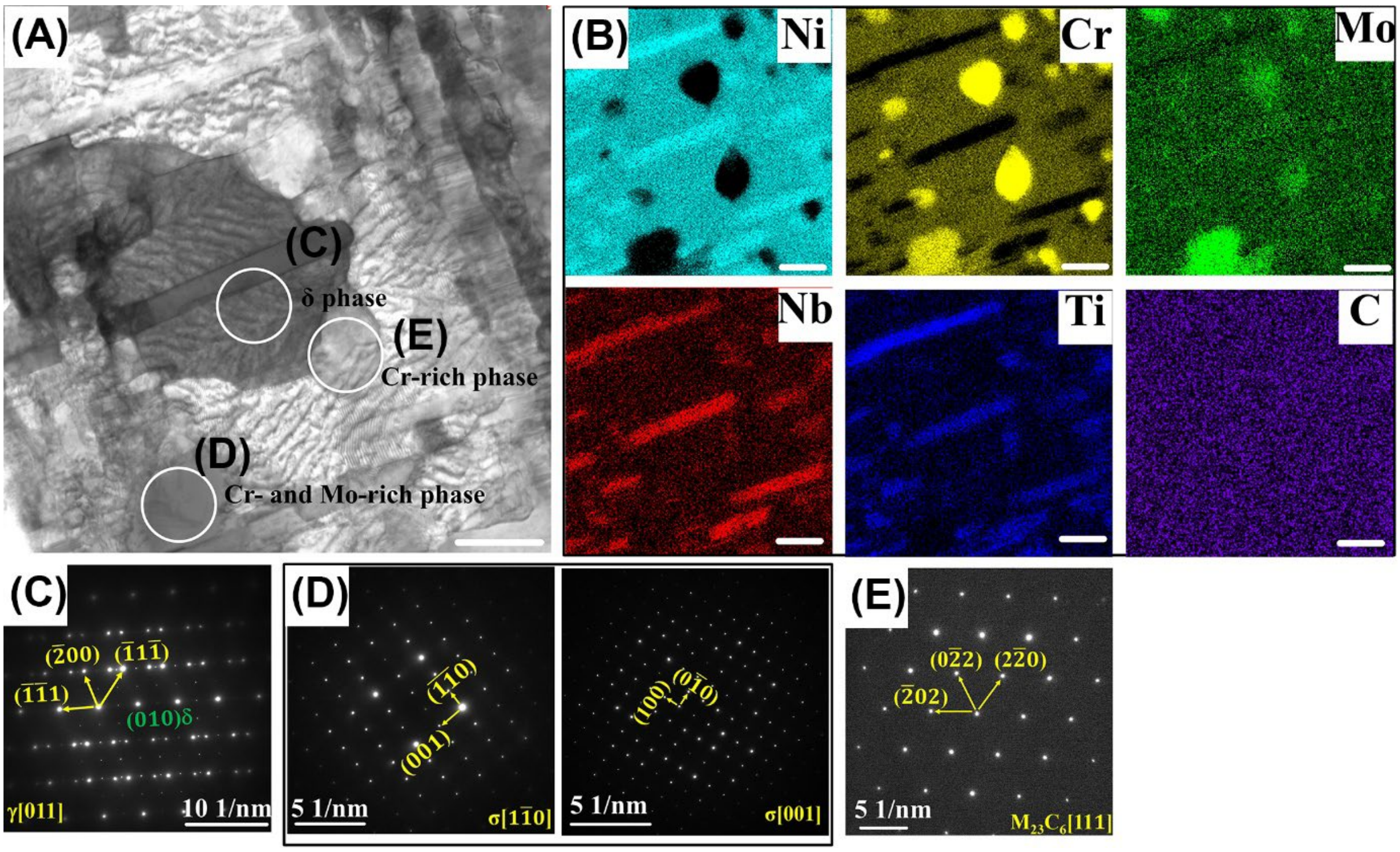

**Fig. S1. (A, B) BF-STEM image of the *Single-aged* sample in the middle region and the corresponding EDS elemental maps of Ni, Cr, Mo, Nb, Ti and C, highlighting the locations of different precipitates. The white scale bars represent 200 nm. (C) SAED pattern acquired from the γ matrix along the [011] zone axis,**

**confirming the presence of δ-phase superlattice reflections with [100] orientation. (D) SAED patterns of the σ-phase acquired along the $[\bar{1}10]$ and [001] zone axes. The characteristic distribution of strong and weak diffraction reflections is consistent with a typical TCP structure. (E) SAED patterns of the carbide precipitates acquired along the [111] zone axe, confirming a typical FCC structure.**

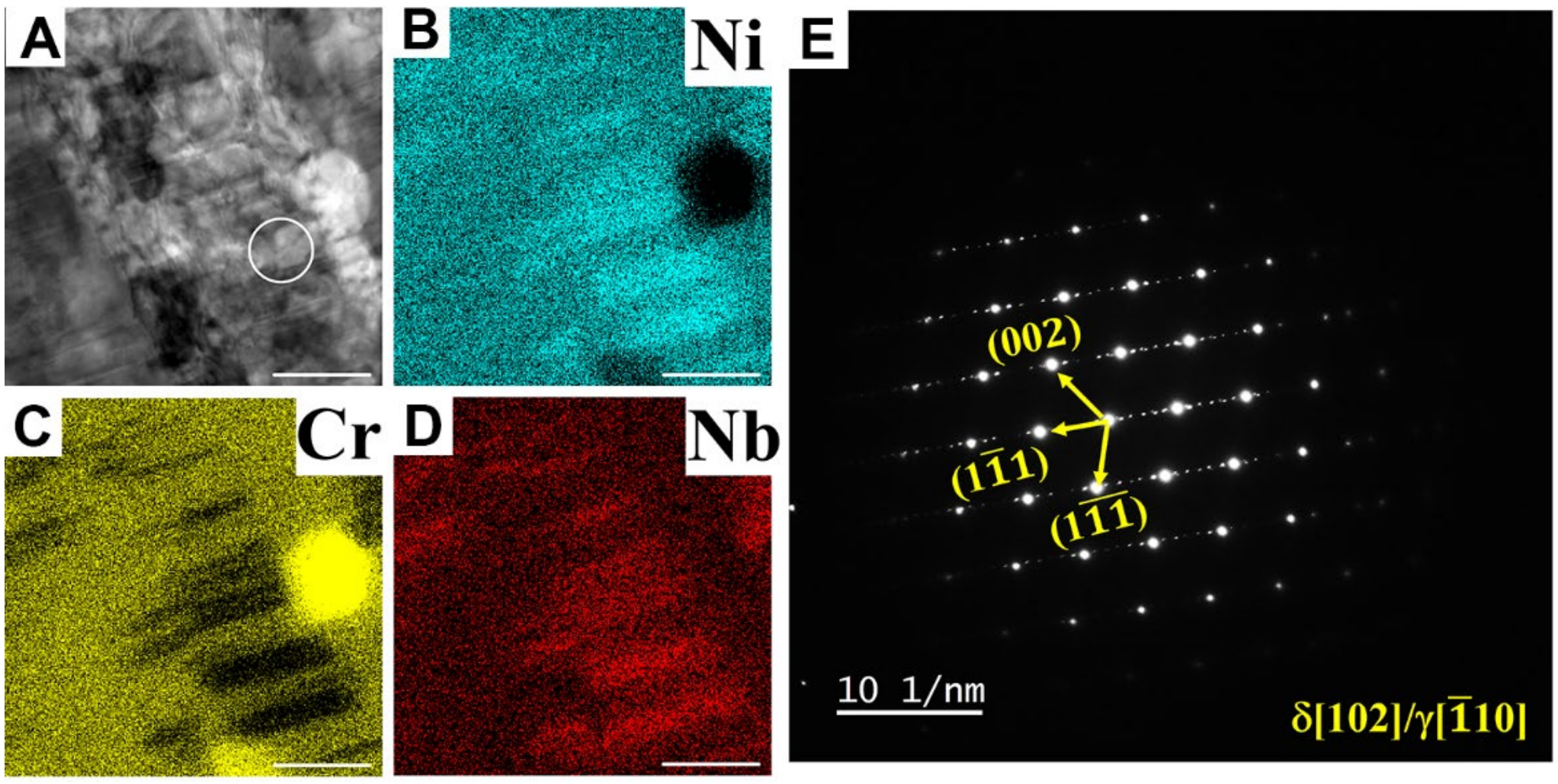

**Fig. S2. (A) BF-STEM image showing the columnar nanotwinned structure region in the *Single-aged sample*. (B-D) Corresponding EDS maps of Ni, Cr, and Nb revealing pronounced elemental segregation associated with δ phase precipitates along twin boundaries. The white scale bars represent 100 nm. (E) SAED pattern acquired from the circled region in (A), indexed as δ[102]/γ[110], demonstrating the crystallographic orientation relationship between the δ phase and the γ matrix.**

The solid-solution strengthening stress, $\sigma_{ss}$, was calculated using the Gypen–Deruyttere model [4], which describes the total contribution of multiple solute elements as follows:

$$\sigma_{ss} = \left[\sum_i \left(a_i \cdot c_i^{1/2}\right)^2\right]^{1/2}$$

where ($a_i$) and ($c_i$) represent the strengthening coefficient and atomic fraction of solute element (i), respectively. The strengthening coefficients of the principal solute elements in Ni were obtained from Ref. [5]. When expressed as hardness, the value was calculated as Hardness=3·$\sigma_{yield}$ [6] expressed in GPa. Based on the γ-matrix compositions listed in Table S1, the solid-solution strengthening contributions of the six kinds of Inconel 725 Alloy samples were estimated.

**Table S1 Compositions*[1] of the γ matrix in Inconel 725 under five heat-treatment conditions (at.%) and the corresponding hardness contributions from solid-solution strengthening.**

| | Ni | Cr | Mo | Fe | Nb | Ti | Al | solid-solution strengthening, GPa |
|---|---|---|---|---|---|---|---|---|
| As-sputtered | 54.9 ± 0.2 | 25.6 ± 0.1 | 4.3 ± 0.3 | 11.8 ± 0.2 | 1.5 ± 0.1 | 1.4 ± 0 | 0.4 ± 0 | 0.98 |
| Single-aged | 56.7 ± 0.2 | 25.4 ± 0.1 | 2.5 ± 0.1 | 13.8 ± 0.1 | 0.6 ± 0 | 0.7 ± 0.1 | 0.4 ± 0.1 | 0.8 |
| Dual-aged | 57.7 ± 0.3 | 24.4 ± 0.2 | 2.5 ± 0.3 | 13.9 ± 0.1 | 0.6 ± 0.1 | 0.6 ± 0 | 0.3 ± 0 | 0.79 |
| Solutionized | 56.2 ± 0.3 | 25.5 ± 0.1 | 3.2 ± 0.2 | 11.9 ± 0.1 | 1.4 ± 0.1 | 1.4 ± 0.1 | 0.4 ± 0.1 | 0.91 |
| Solutionized+single-aged sample | 55.9 ± 0.3 | 25.9 ± 0.6 | 3.0 ± 0.2 | 12.3 ± 0.5 | 1.3 ± 0.4 | 1.2 ± 0.3 | 0.3 ± 0.2 | 0.89 |
| Solutionized+dual-aged sample | 56.3 ± 1.1 | 25.5 ± 1.2 | 3.2 ± 0.2 | 12.4 ± 0.5 | 1.3 ± 0.5 | 1.1 ± 0.3 | 0.3 ± 0.1 | 0.9 |

***[1]:The γ-matrix compositions were determined from multiple STEM-EDS point analyses and are reported as the mean ± standard deviation. The corresponding hardness contributions from solid-solution strengthening were estimated using the Gypen–Deruyttere model.**

**Table S2 Compositions*2 of the δ phase in Inconel 725 under five heat-treatment conditions (wt.%).**

| | | Ni | Cr | Mo | Fe | Nb | Ti | Al |
|---|---|---|---|---|---|---|---|---|
| Single-aged | Top | 78.3 ± 1.7 | 8.6 ± 2.2 | 0.2 ± 0.3 | 6.0 ± 1.2 | 3.2 ± 1.4 | 3.5 ± 0.7 | 0.2 ± 0.2 |
| | middle | 73.9 ± 0.7 | 6.8 ± 1.9 | 3.6 ± 0.7 | 5.5 ± 0.9 | 6.1 ± 1.1 | 3.5 ± 0.6 | 0.6 ± 0.2 |
| | Bottom | 74.8 ± 0.9 | 7.9 ± 2.7 | 3.3 ± 2.2 | 5.8 ± 2.2 | 6.6 ± 3.1 | 1.6 ± 0.6 | 0 |
| Dual-aged | Top | 71.2 ± 1.3 | 9.5 ± 0.3 | 4.3 ± 2.0 | 4.4 ± 0.1 | 8.5 ± 1.0 | 2.0 ± 0.3 | 0.1 ± 0.1 |
| | middle | 72.3 ± 1.8 | 7.8 ± 0.7 | 2.5 ± 0.5 | 7.0 ± 0.1 | 7.8 ± 1.3 | 2.5 ± 0 | 0.1 ± 0.1 |
| | Bottom | 76.9 ± 0.9 | 2.2 ± 0.8 | 0.5 ± 1.1 | 1.9 ± 0.9 | 15.5 ± 0.4 | 2.8 ± 0.5 | 0.1 ± 0.1 |
| Solutionized | Top | 63.5 ± 0.8 | 6.3 ± 1.1 | 4.9 ± 0.3 | 6.8 ± 0.7 | 15.4 ± 0.2 | 2.9 ± 0.2 | 0.2 ± 0.1 |
| | middle | 77.9 ±10.2 | 3.8 ± 1.8 | 4.4 ± 5.9 | 3.1 ± 0.9 | 8.9 ± 7.6 | 1.9 ± 1.4 | 0 |
| | Bottom | 75.7 ± 0.9 | 5.9 ± 0.8 | 2.4 ± 0.5 | 3.5 ± 0.5 | 9.6 ± 0.9 | 2.7 ± 0.1 | 0.2 ± 0.1 |
| Solutionized+single-aged sample | Top | 67.4 ± 4.4 | 6.5 ± 1.2 | 0.3 ± 0.2 | 7.2 ± 2.5 | 14.5 ± 0.2 | 3.9 ± 0.9 | 0.2 ± 0.2 |
| | middle | 63.9 ± 4.7 | 6.9 ± 1.4 | 11.4 ± 2.5 | 5.2 ± 2.0 | 9.6 ± 2.5 | 2.8 ± 0.3 | 0.2 ± 0.2 |
| | Bottom | 65.0 ± 1.1 | 13.4 ± 0.0 | 3.7 ± 0.8 | 3.0 ± 0.3 | 13.9 ± 0.1 | 0.8 ± 0.0 | 0.2 ± 0.1 |
| Solutionized+dual-aged sample | Top | 65.2 ± 2.1 | 4.5 ± 1.3 | 6.1 ± 0.6 | 3.4 ± 0.2 | 18.2 ± 2.6 | 2.5 ± 0.3 | 0.2 ± 0.1 |
| | middle | 68.9 ± 2.9 | 2.4 ± 0.6 | 2.8 ± 2.3 | 2.0 ± 0.2 | 20.6 ± 1.4 | 3.3 ± 0.3 | 0 |
| | Bottom | 72.9 ± 2.2 | 4.0 ± 1.7 | 0.4 ± 0.8 | 3.9 ± 0.9 | 12.8 ± 1.1 | 5.5 ± 0.4 | 0.5 ± 0.1 |

**Table S3 Compositions*2 of the Cr-rich precipitates in Inconel 725 under five heat-treatment conditions (wt.%).**

| | | Ni | Cr | Mo | Fe | Nb | Ti | Al |
|---|---|---|---|---|---|---|---|---|
| Single-aged | Top | 3.1 ± 1.1 | 80.9 ± 2.1 | 12.1 ± 1.7 | 3.1 ± 0.3 | 0.6 ± 0.6 | 0.1 ± 0.1 | 0.1 ± 0.1 |
| | middle | 27.7 ± 1.5 | 35.9 ± 1.0 | 26.1 ± 2.2 | 9.5 ± 1.0 | 0.5 ± 0.7 | 0.2 ± 0.2 | 0.1 ± 0.1 |
| | Bottom | 28.7 ± 1.1 | 37.0 ± 1.2 | 23.6 ± 1.9 | 10.4 ± 0.7 | 0.1 ± 0.3 | 0.1 ± 0.2 | 0.1 ± 0.1 |
| Dual-aged | Top | 3.8 ± 2.5 | 75.7 ± 8.1 | 13.3 ± 6.6 | 3.2 ± 2.3 | 3.2 ± 3.4 | 0.5 ± 0.7 | 0.3 ± 0.4 |
| | middle | 25.9 ± 2.0 | 39.6 ± 3.0 | 23.2 ± 3.3 | 10.2 ± 1.3 | 0.5 ± 0.1 | 0.3 ± 0.3 | 0.4 ± 0.4 |
| | Bottom | 28.2 ± 1.8 | 37.8 ± 2.2 | 23.4 ± 3.1 | 9.4 ± 1.7 | 0.9 ± 0.7 | 0.3 ±0.3 | 0 |
| Solutionized | Top | 33.9 ± 2.3 | 32.6 ± 1.8 | 18.7 ± 2.3 | 9.1 ± 1.4 | 3.9 ± 1.6 | 0.8 ± 0.3 | 0.9 ± 0.4 |
| | middle | 29.4 ± 7.4 | 34.5 ± 7.2 | 24.8 ±10.9 | 8.8 ± 3.6 | 1.6 ±2.9 | 0.6 ± 0.7 | 0.3 ± 0.5 |
| | Bottom | 30.8 ± 1.1 | 34.3 ± 1.0 | 24.3 ± 1.8 | 9.0 ± 0.4 | 0.9 ± 0.4 | 0.5 ± 0.1 | 0.2 ± 0.1 |
| Solutionized+single-aged sample | Top | 30.5 ± 6.2 | 35.2 ± 6.2 | 21.5 ± 7.5 | 9.2 ± 2.7 | 2.6 ± 3.6 | 0.6 ± 0.8 | 0.3 ± 0.4 |
| | middle | 34.2 ± 8.0 | 34.0 ± 5.0 | 20.1 ± 8.3 | 9.0 ± 2.8 | 1.9 ± 2.7 | 0.5 ± 0.7 | 0.3 ± 0.3 |
| | Bottom | 28.6 ± 3.7 | 35.0 ± 5.0 | 24.6 ± 6.9 | 8.8 ± 1.8 | 2.3 ± 2.4 | 0.4 ± 0.5 | 0.2 ± 0.3 |
| Solutionized+dual-aged sample | Top | 26.1 ± 1.6 | 33.5 ± 2.3 | 29.6 ± 2.6 | 8.5 ± 0.8 | 2.0 ± 1.5 | 0.2 ± 0.2 | 0.1 ± 0.1 |
| | middle | 25.8 ± 1.1 | 34.0 ± 1.6 | 29.4 ±2.3 | 8.4 ± 0.8 | 2.2 ± 1.4 | 0.2 ± 0.2 | 0.1 ± 0.1 |
| | Bottom | 25.7 ± 1.2 | 36.9 ±2.4 | 26.8 ± 2.0 | 8.9 ± 0.7 | 1.2 ± 1.2 | 0.3 ± 0.3 | 0.1 ± 0.1 |

**Table S4 Compositions*2 of the γ′/γ″ precipitates in the Solutionized+single-aged and Solutionized+dual-aged sample (wt.%).**

| | | Ni | Cr | Mo | Fe | Nb | Ti | Al |
|---|---|---|---|---|---|---|---|---|
| Solutionized+single-aged sample | | 56.4 ± 5.6 | 17.5 ± 2.6 | 8.1 ± 2.6 | 7.9 ± 3.1 | 6.7 ± 2.0 | 2.6 ± 1.1 | 0.8 ± 0.5 |
| Solutionized+dual-aged sample | | 58.0 ± 2.2 | 13.2 ± 1.4 | 8.0 ± 1.2 | 6.5 ± 0.8 | 10.7 ± 1.2 | 2.7 ± 0.7 | 0.9 ± 0.3 |

**[*2]:The compositions of δ phase, Cr-rich precipitates, and γ′/γ″ precipitates were determined from multiple STEM-EDS map analyses and are reported as the mean ± standard deviation.**